\documentclass{article}
\usepackage{bris}
\usepackage{cite}
\usepackage{epsfig}

\def\Journal#1#2#3#4{{#1} {\bf #2}, #3 (#4)}

\def\PLB{{\em Phys. Lett.}  B}

\def\EUR{{\em Eur. Phys.} C}

\def\st{\scriptstyle}

\def\ra{\rightarrow}

\def\be{\begin{equation}}
\def\ee{\end{equation}}
\def\bea{\begin{eqnarray}}
\def\eea{\end{eqnarray}}

\begin{document}

\begin{titlepage}{BRIS/HEP/2000--04}{July 2000}
\title{STRUCTURE FUNCTIONS AT LOW X AND Q$^2$ AT ZEUS}

\author{J. P. SCOTT\\
(On behalf of the ZEUS collaboration)}

\begin{abstract} 
Using 3.9 $pb^{-1}$ of data from ep $\ra$ eX interactions recorded using the ZEUS detector in 1997,
the proton structure function,  F$_2$, has been measured in the range 0.015 GeV$^2$ $<$ Q$^2$ 
$<$ 0.65 GeV$^2$
and $6 \times 10^{-7}$ $<$ x $<$ $1 \times 10^{-3}$. The analysis is based on data taken incorporating new detector components. Compared with the previous
analysis, these components allow improved background suppression and better control of systematic uncertainties, extending the accessible kinematic region towards lower Q$^2$.
\end{abstract}
\end{titlepage}

\section{Introduction}
The measurement of the proton struture function F$_2$ at low Q$^2$ has been used recently by ZEUS \cite{zeusfit} to study the transition between photoproduction and deep inelastic scattering. The first measurement by ZEUS \cite{f2low} using 1995 data has now been followed by a higher precision measurement using dedicated detector components.

A measurement of F$_2$ has been made in the range 0.015 GeV$^2$ $<$ Q$^2$
 $<$ 0.65 GeV$^2$ and $6 \times 10^{-7}$ $<$ x $<$ $1 \times 10{^-3}$, where Q$^2$ is the four-momentum transfer squared and x is the Bjorken scaling variable. The measurement was made using 3.9 pb$^-1$ of data taken using special triggers in 1997 and, compared to the previous result, covers a larger kinematic region with improved accuracy.

\section{Analysis} 
Reconstruction of the scattered positron is performed using dedicated detector compontents: the Beam Pipe Calorimeter (BPC) and the Beam Pipe Tracker (BPT) of the ZEUS detector.

The BPC is a small Tungsten-scintillator sampling calorimeter that detects
positrons scattered at angles of 18-32mrad with respect to the positron beam direction. It has an energy resolution of $\sigma_E$ = 0.17$/\sqrt{E[GeV]}$.

The BPT consisted of two silicon microstrip detectors. A track is reconstructed
as the straight line between a hit on each detector and provides information on the positron scattering angle and
 impact point on the BPC. This helps eliminate background and reduce the systematic uncertainties.

\section{Results}
Figure 1 shows F$_2$ as a function of x in bins of Q$^2$. Down to Q$^2$$\approx$1 the data is well described by NLO QCD fits. The steep rise in F$_2$ at low x observed at higher Q$^2$ persists down to the low Q$^2$ region. This rise though, becomes shallower as Q$^2$ decreases into the new region of measuremement. At low values of Q$^2$ the data are well described by Regge theory.

\section{Conclusions}
The ZEUS collaboration has measured F$_2$ in the region  0.015 GeV$^2$ $<$ Q$^2$
 $<$ 0.65 GeV$^2$ and $6 \times 10^{-7}$ $<$ x $<$ $1 \times 10^{-3}$ with high precision. The data can be used to examine the transition between perturbative and non-perturbative QCD.

\begin{figure}[t]
\begin{center}
\epsfig{file=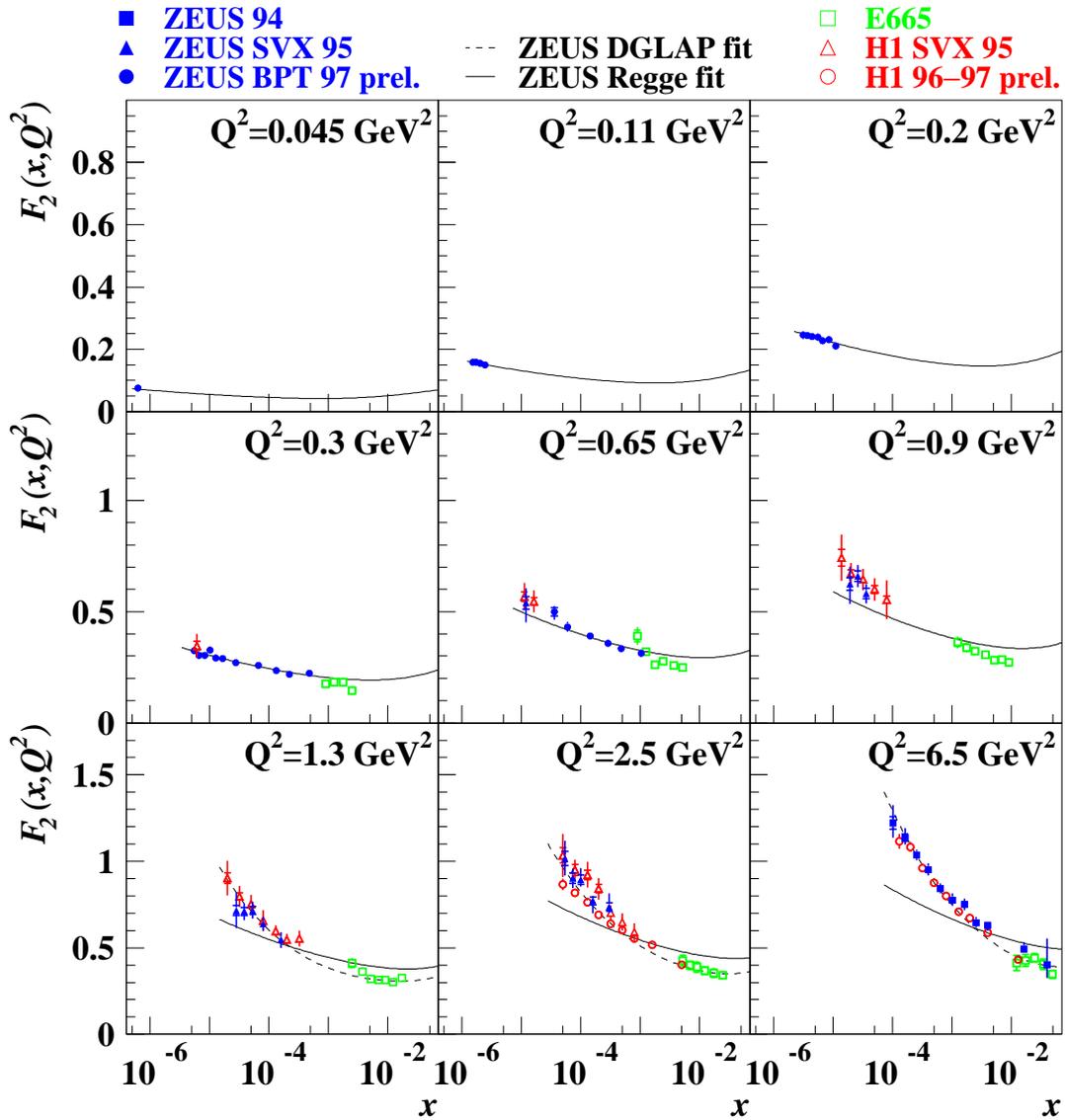,height=15cm}
\caption{F$_2$ versus x in bins of Q$^2$, compared to other measurements
from ZEUS, H1 and fixed target experiments.
\label{fig:radish}}
\end{center}
\end{figure}

\end{document}